\documentclass{JINST}

\newcommand{\gx}{\textsc{GlueX}}

\usepackage{amsmath}

\title{The \gx{} DIRC Project}

\author{J.~Stevens$^a$\thanks{Corresponding author}, F.~Barbosa$^a$, J.~Bessuille$^c$, E.~Chudakov$^a$, R.~Dzhygadlo$^b$, C.~Fanelli$^c$, J.~Frye$^d$, J.~Hardin$^c$, J.~Kelsey$^c$, M.~Patsyuk$^c$, C.~Schwartz$^b$, J.~Schwiening$^b$, M.~Shepherd$^d$, T.~Whitlatch$^a$ and M.~Williams$^c$ \\
\llap{$^a$}Thomas Jefferson National Accelerator Facility, \\ Newport News, VA, United States\\
\llap{$^b$} GSI Helmholtzzentrum f\"{u}r Schwerionenforschung GmbH,\\ Darmstadt, Germany \\
\llap{$^c$} Massachusetts Institute of Technology, \\ Cambridge, MA, United States\\
\llap{$^d$} Indiana University, \\ Bloomington, IN, United States\\

  E-mail: \email{jrsteven@jlab.org}}

\abstract{ 

The \gx{} experiment was designed to search for and study the pattern of gluonic excitations in the meson spectrum produced through photoproduction reactions at a new tagged photon beam facility in Hall D at Jefferson Laboratory.  The particle identification capabilities of the \gx{} experiment will be enhanced by constructing a DIRC (Detection of Internally Reflected Cherenkov light) detector, utilizing components of the decommissioned BaBar DIRC.  The DIRC will allow systematic studies of kaon final states that are essential for inferring the quark flavor content of both hybrid and conventional mesons.  The design for the \gx{} DIRC is presented, including the new expansion volumes that are currently under development.

 }

\begin{document}

\section{Introduction}
\label{Sec:Intro}

\begin{figure}
    \begin{center}
        \includegraphics[width=0.9\textwidth]{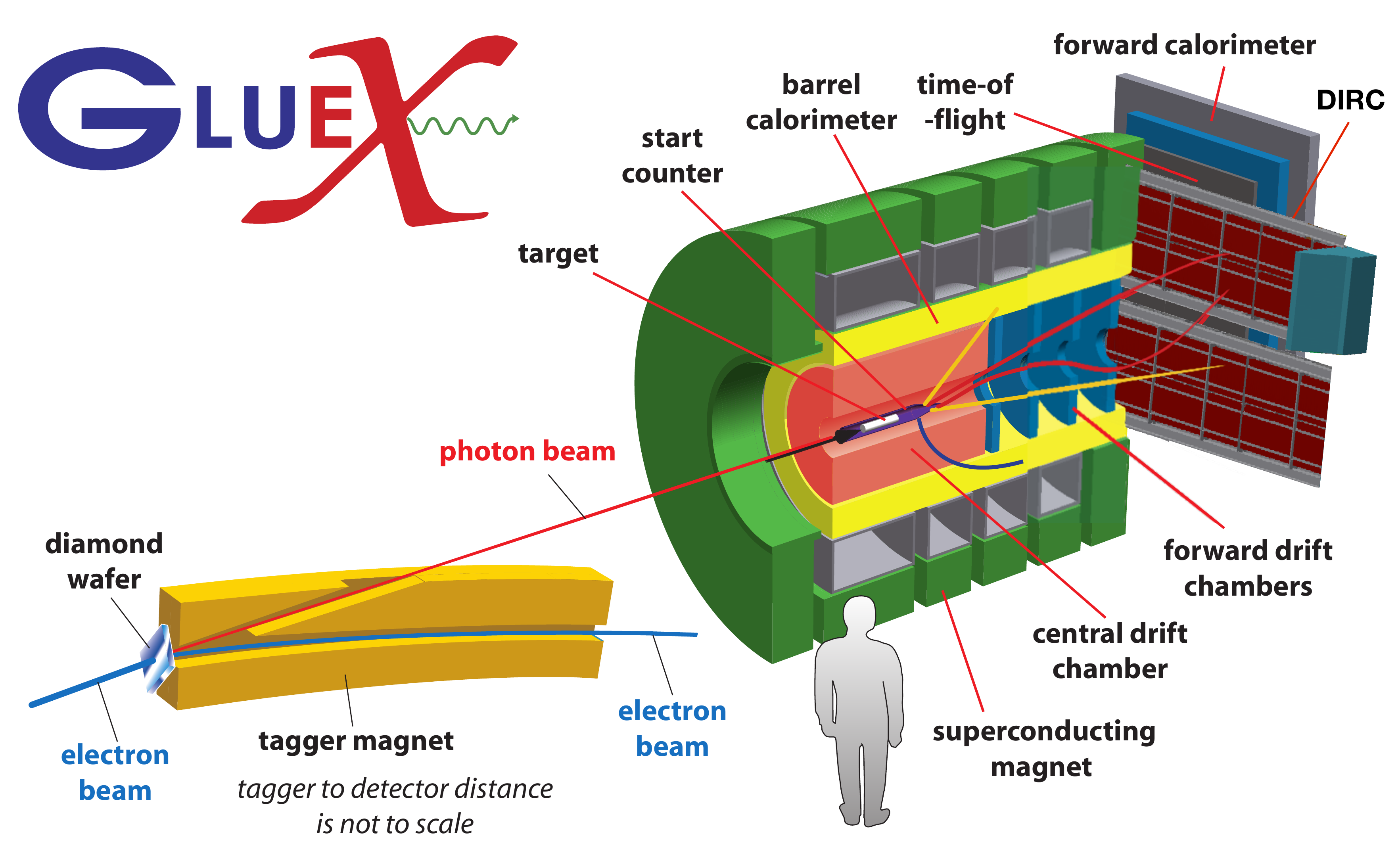}
    \caption{A schematic of the new Hall D beamline and \gx{} detector at Jefferson Laboratory.  The DIRC detector will be installed directly upstream of the time-of-flight detector in the forward region.}
\label{fig:GlueXcartoon}
\end{center}
\end{figure}

Jefferson Laboratory is nearing the completion of an upgrade to double the beam energy of the Continuous Electron Beam Accelerator Facility (CEBAF) and significantly enhance the experimental facilities, including an entirely new experimental area, Hall D.  The \gx{} experiment, located in Hall D and shown schematically in Fig.~\ref{fig:GlueXcartoon}, utilizes a tagged photon beam derived from the 12 GeV electron beam's coherent bremsstrahlung radiation from a thin diamond wafer.  The primary goal of the experiment is to search for and ultimately study an unconventional class of mesons, known as hybrids mesons, which contain an intrinsic gluonic component to their wave functions~\cite{PAC30,PAC40,PAC42}.  These hybrid meson states are predicted by Lattice QCD calculations~\cite{Dudek:2013yja}, and provide an opportunity to quantitatively test our understanding of the strong nuclear force in this non-perturbative regime.

\begin{figure}
    \begin{center}
        \includegraphics[width=0.49\textwidth]{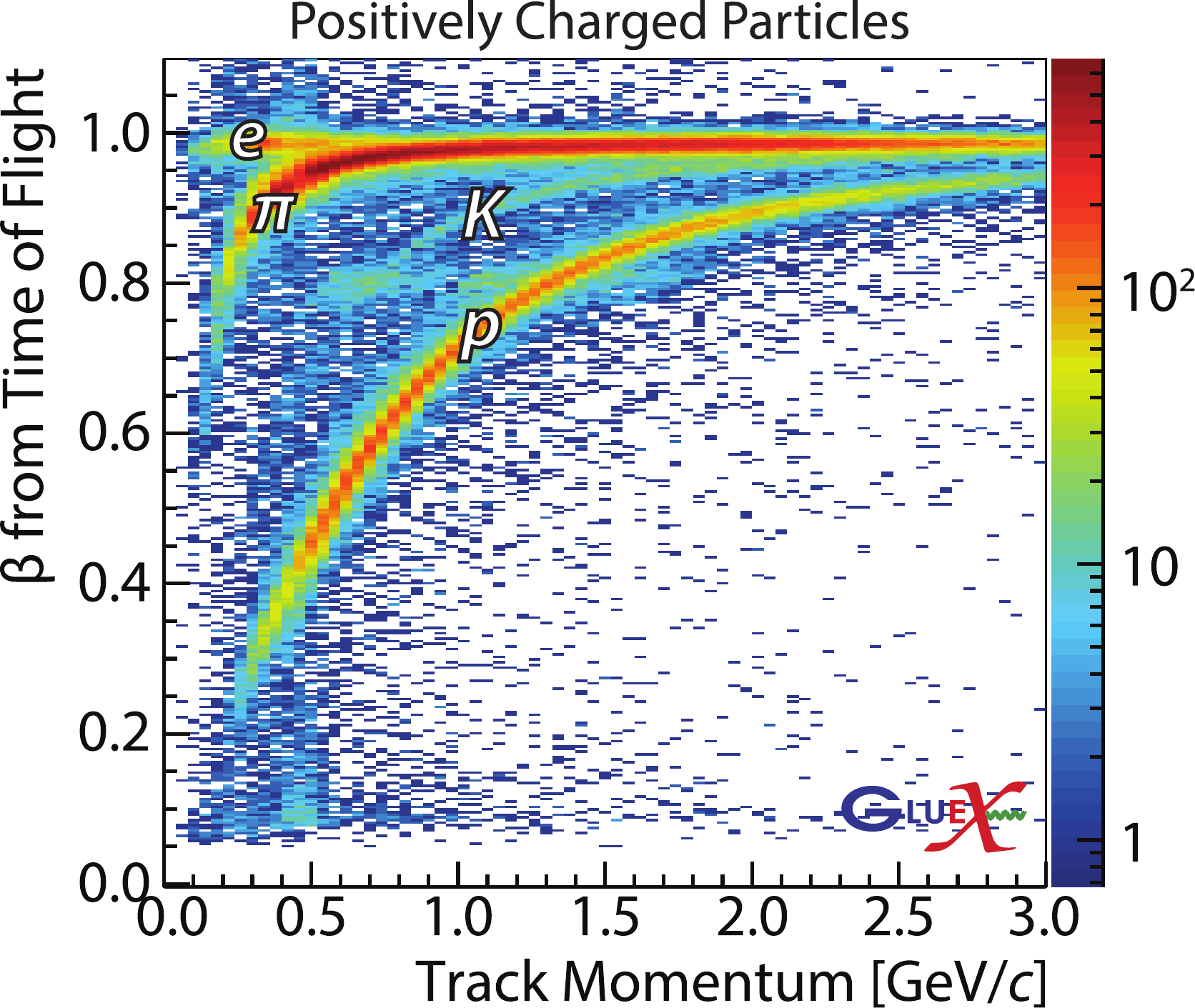}
        \includegraphics[width=0.49\textwidth]{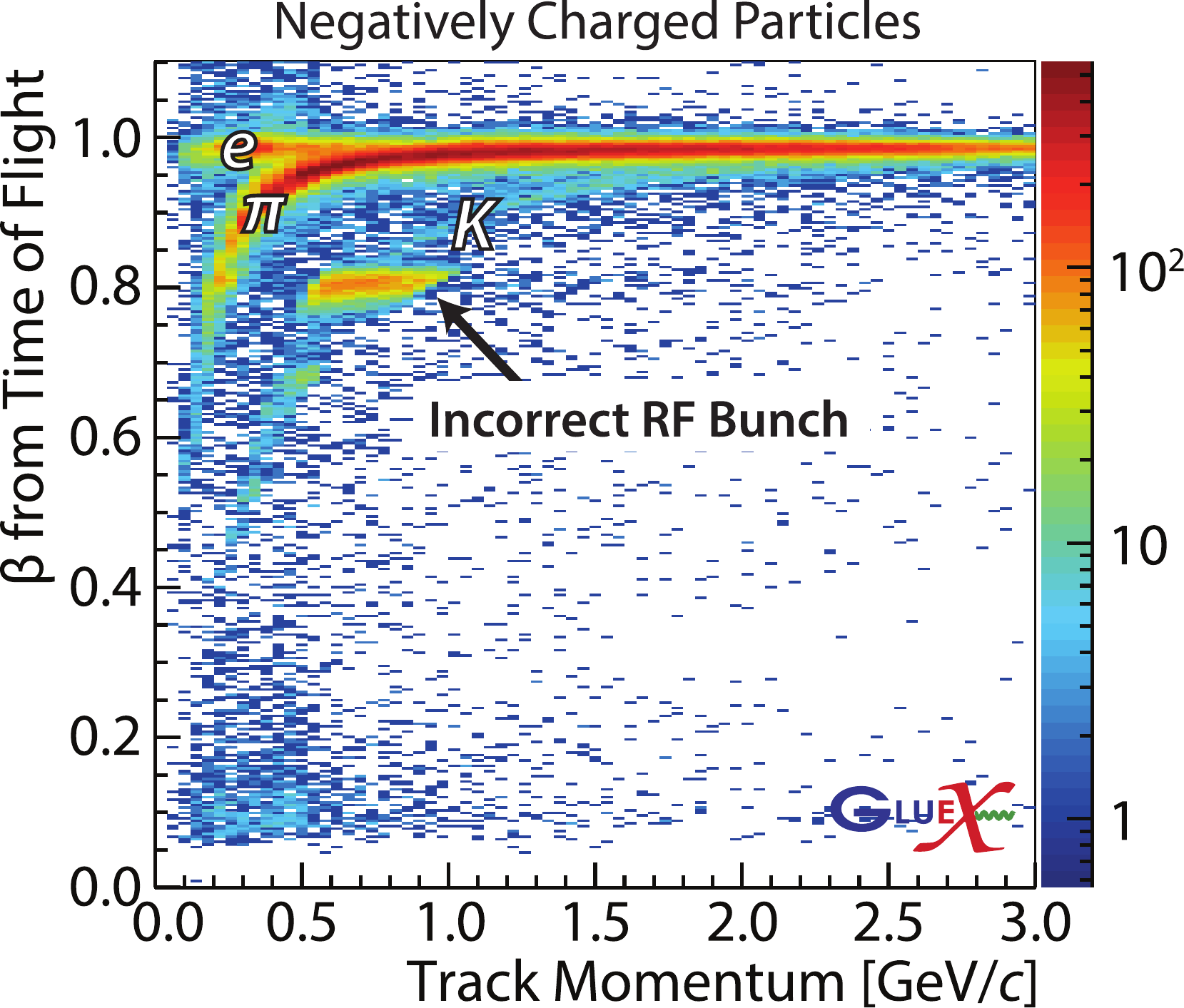}
    \caption{The charged particle $\beta$ as determined from both time-of-flight and path-length measurements in the detector versus the reconstructed particle momentum, for positively charged particles (left)  and negatively charged particles (right).}
\label{fig:tof}
\end{center}
\end{figure}

Construction and installation of the baseline \gx{} detector was completed in 2014 and since that time an extensive commissioning program has been carried out in Hall D, with first photon beams delivered in Fall 2014 and continuing through the final commissioning run in Spring 2016~\cite{Ghoul:2015ifw}.  The detector systems have been commissioned and initial calibrations are in advanced stages.  The particle identification (PID) capabilities of the \gx{} detector have been studied with this initial data, and in the forward region the time-of-flight (TOF) detector performance has approached its design specifications for providing $\pi/K$ separation up to $p\sim2$~GeV$/c$, as can be seen in Fig.~\ref{fig:tof}.  An initial physics program to search for and study hybrid mesons which decay to non-strange final state particles will begin in Fall 2016, however an upgrade to the PID capabilities is needed to fully exploit the discovery potential of the \gx{} experiment by studying the quark flavor content of the potential hybrid states.

\begin{figure}
    \begin{center}
        \includegraphics[width=0.85\textwidth]{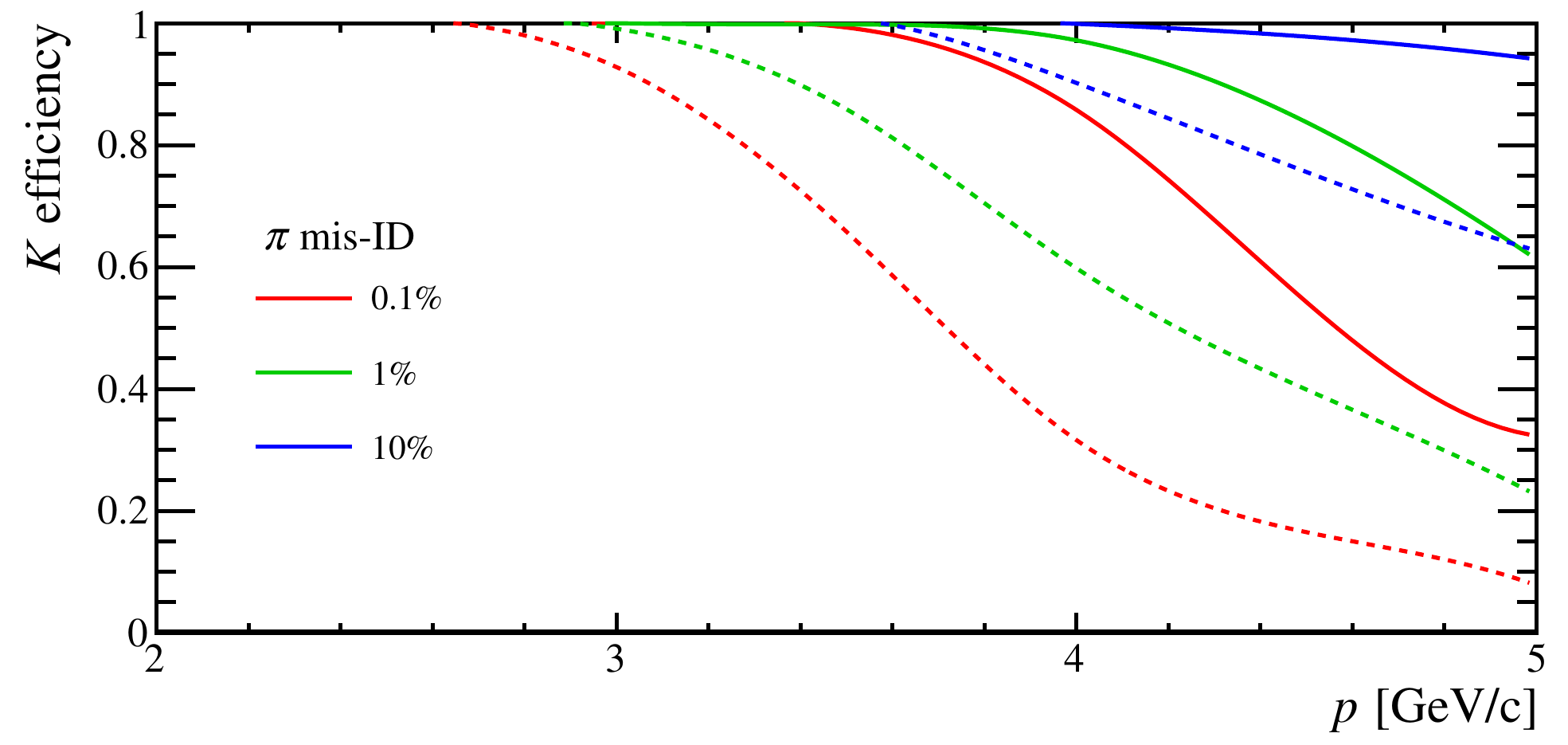}
    \caption{Kaon efficiency {\em vs} momentum for pion mis-identification probabilities of 0.1, 1, and 10\%, where pion mis-identification probability is defined as the probability for a given charged pion track to be incorrectly identified as a charged kaon.  The dashed curves show the conservative performance assumed in a simplified model used to evaluate the impact of a DIRC on particular physics channels in Ref.~\cite{PAC42}, while the solid curves show the improved performance achieved in simulation for the current design.}
\label{fig:performance}
\end{center}
\end{figure}

The proposed PID upgrade for \gx{} will utilize one-third of the fused silica radiators from the BaBar DIRC (Detection of Internally Reflected Cherenkov light) detector~\cite{Adam:2004fq}, with two new, compact expansion volumes.  The \gx{} DIRC design is described in detail in Ref.~\cite{dirc_tdr} and summarized in the next section.  The expected performance of this system was evaluated through extensive simulation studies, and can be seen in Fig.~\ref{fig:performance} in terms of kaon efficiency for different pion mis-identification probabilities.  Several potential hybrid meson decay modes containing charged kaons were studied within the \gx{} analysis framework, assuming the  conservative performance given by the dashed curves in Fig.~\ref{fig:performance}, which already yielded significantly increased efficiency and purity for these decay modes~\cite{PAC42}.  Since then, extensive simulation studies have shown that the current design performance for the \gx{} DIRC, given by the solid curves in Fig.~\ref{fig:performance}, exceed those initial assumptions and will significantly extend the physics reach of the experiment.  

\section{\gx{} DIRC Design}
\label{Sec:Design}

As shown in Fig.~\ref{fig:GlueXcartoon}, the fused silica radiators for the \gx{} DIRC will be oriented in a plane normal to the beam covering the forward region of the spectrometer, where the PID requirements are most stringent due to the higher momentum of the particles at these small angles.  The acceptance in this region of the spectrometer is limited by the solenoid to $\theta \lesssim 11^{\circ}$, and requires four BaBar boxes, each containing 12 fused silica bars, to fully cover the acceptance.  These ``bar boxes" from BaBar are $\sim$5 m long and will be oriented horizontally in Hall D directly upstream of the TOF, with two bar boxes both above and below the beam.  Each pair of the un-modified bar boxes will be coupled to a new, compact expansion volume, referred to as the optical box.  One optical box will be to beam left and the other to beam right to avoid vertical interference between the optical boxes.  Additionally the support structure for the DIRC will allow the pairs of bar boxes to move vertically out the active area of the detector for experiments which require minimal material in front of the forward calorimeter.

\subsection{Expansion Volume}

The initial photon camera of the BaBar DIRC detector was very large and filled with 6000 liters of purified water~\cite{Adam:2004fq}.  The design was updated for the SuperB\footnote{SuperB was designed to be the successor to BaBar but was canceled in 2012.} experiment to include focusing mirrors that permit detecting the Cherenkov light produced in the quartz radiator using a much more compact design~\cite{Dey:2014fda}, therefore minimizing backgrounds and reducing costs by limiting the area of the photodetector plane.  This focusing DIRC (FDIRC) was designed and prototyped at SLAC with the constraint that the BaBar bar boxes could not be altered, which is also a requirement for the \gx{} DIRC.  Therefore, for the \gx{} DIRC optical box design we, quite naturally, began with the FDIRC prototype design, with some modifications allowed by our unique geometry.

\begin{figure}
    \begin{center}
        \includegraphics[width=0.6\textwidth]{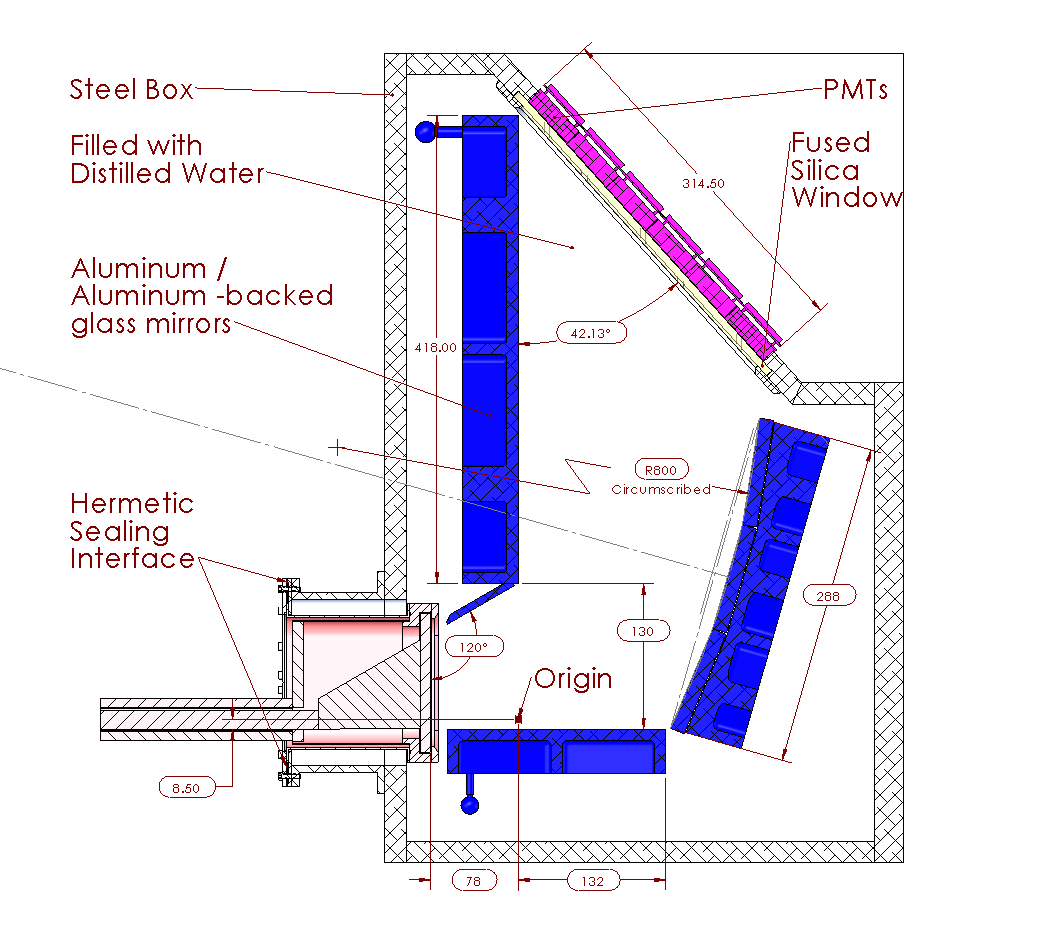}
        \includegraphics[width=0.37\textwidth]{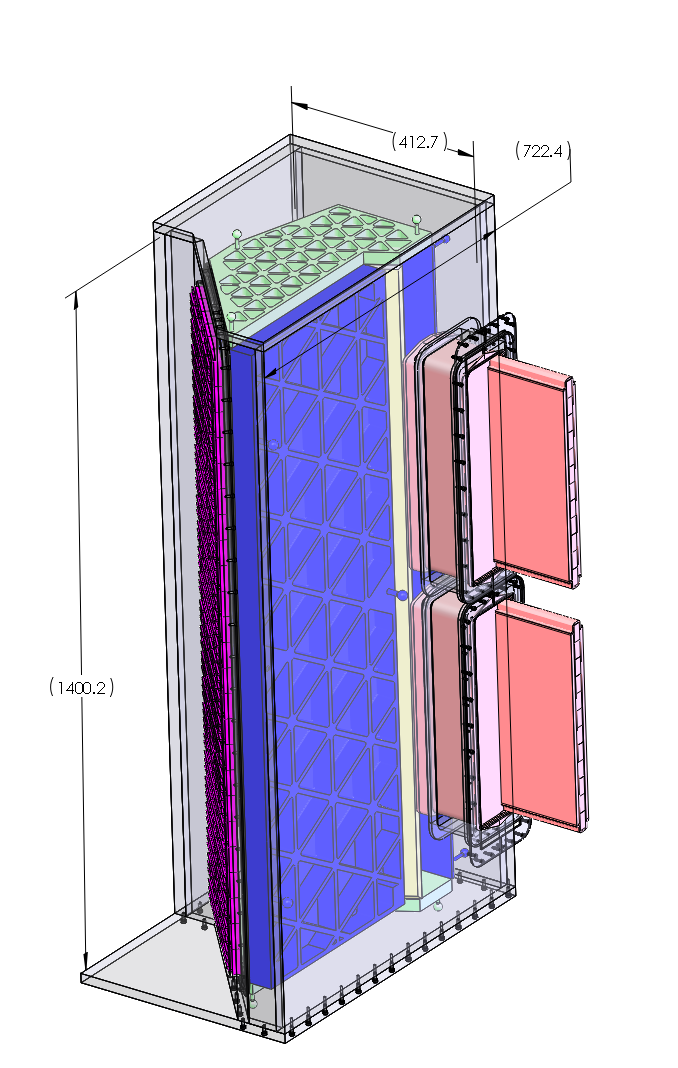}
    \caption{ Schematic diagram of the optical box mirror geometry (left) and complete assembly, with mirror mounts, {\em etc.}, coupled to two of the bar boxes from BaBar (right). }
\label{fig:box}
\end{center}
\end{figure}

\begin{figure}
    \begin{center}
        \includegraphics[width=0.49\textwidth]{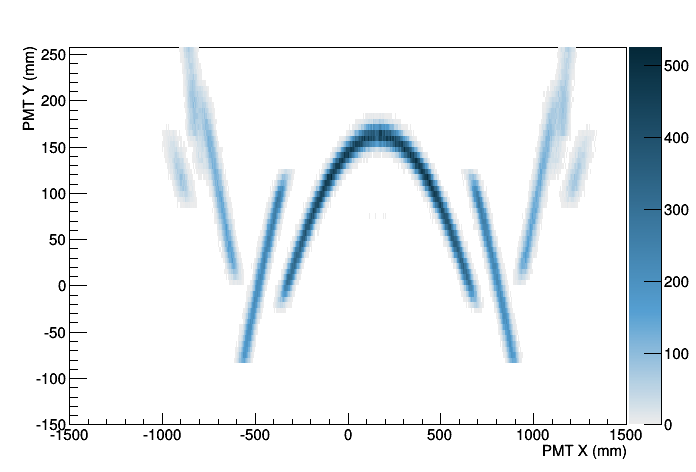}
        \includegraphics[width=0.49\textwidth]{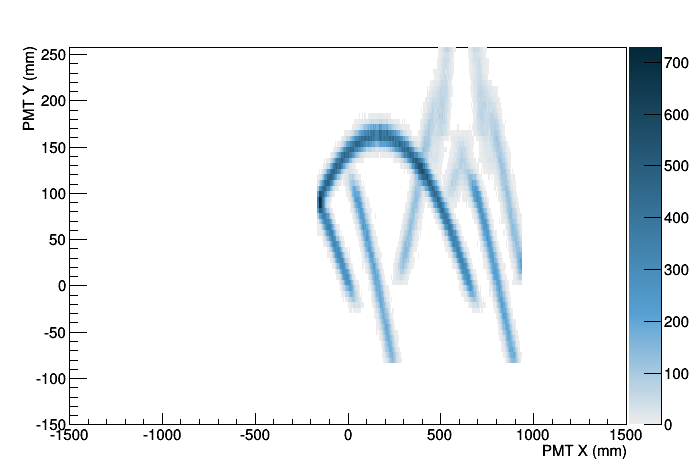}
    \caption{ PMT hit patterns for charged particles incident perpendicular to the plane defined by the fused silica radiators, with an optical box containing three flat mirrors approximating a cylindrical radius for an optical box with an arbitrary vertical length (left) and the current design of a $\sim 1.4$ m long optical box with side mirrors (right), shown schematically in Fig. 4.}
\label{fig:pmtplane}
\end{center}
\end{figure}

Figure~\ref{fig:box} (left) shows a schematic for the \gx{} DIRC optical box.  The volume is filled with distilled water, whose index of refraction is similar to the fused silica radiators, and contains multiple flat mirrors which direct the light to the photodetector plane that interfaces with the water volume via a fused silica window.  The SuperB FDIRC design utilized a cylindrical mirror to focus the Cherenkov light onto the PMT plane.  We have replaced this with a three-segment approximation of the cylindrical curvature, using only flat mirrors for simplicity in construction and alignment.  Figure~\ref{fig:pmtplane} (right) shows the occupancy in the photodetector plane for charged particles incident perpendicular to the plane defined by the fused silica radiators, with an optical box of arbitrarily vertical length to show the full hit pattern.  The separated segments of the Cherenkov ``ring" shown there are due to the three-segment mirror, which discretizes the hit pattern.  Simulation studies have shown that this discretization due to the three-segment mirror does not adversely effect the PID performance of the detector~\cite{dirc_tdr}

The planar alignment of the fused silica radiators in \gx{} also permits using a large common readout system for multiple bar boxes, as opposed to the barrel design of SuperB which required a single, small expansion volume for each bar box which can lead to ambiguities in the reconstruction.  As shown in Fig.~\ref{fig:box} (right), two bar boxes are coupled to each optical box with a vertical length of $\sim 1.4$ m and side mirrors to contain the Cherenkov light near the ends of the optical box.  An optical box of arbitrary vertical length would provide the entire hit pattern to be studied without reflections, see Fig.~\ref{fig:pmtplane} (left).  With the $\sim 1.4$ m optical box length in \gx{}, Fig.~\ref{fig:pmtplane} (right), reflections from the side mirrors do introduce some folding of the hit pattern on the photodetector plane occurs, but these can be resolved by utilizing timing information in the reconstruction algorithms.  

\subsection{Simulation and Reconstruction}

In order to optimize the design, we need to be able to convert the complicated patterns observed on the photodetector plane into a Cherenkov angle $\theta_C$.  The BaBar experiment used a so-called look up table (LUT) approach to determining the $\theta_C$~\cite{Adam:2004fq}.  This method generates a large number of photons at various Cherenkov angles exiting each bar, then tracks them to the PMT plane.  For each hit recorded in a PMT pixel, the LUT provides a list of all possible photon propagation vectors that could have led to a photon hitting the pixel.  Plotting the cumulative distribution of possible $\theta_C$ values for all photons associated to a charged particle provides a means for determining the actual value of $\theta_C$.  This algorithm has been implemented by the PANDA Barrel DIRC collaboration~\cite{Dzhygadlo:2016uqi} and has been adapted for {\sc Geant4} MC in the \gx{} DIRC geometry. 

An alternative approach is to obtain a probability density function (PDF) for the distribution of photons expected on the photodetector plane for each charged-particle hypothesis.  Using this PDF and the PMT hits associated to the particle, the likelihood is easily obtained for each hypothesis.  Arbitration between particle types is then performed using the likelihood ratio, or equivalently the difference in the log of the likelihoods, formed from two hypotheses.  The patterns of photon hits produced by the \gx{} DIRC are complicated and, as of yet, we do not know how to determine them analytically.  Instead, we numerically determine the PDFs using Kernel Density Estimation (KDE).  The KDE method constructs an estimate for the PDF for a single charged particle as follows:
\begin{itemize}
\item a large number, $\mathcal{O}(10^5)$, of photons are randomly generated for the particle under a single hypothesis;
\item each MC photon is traced to the photodetector plane (this is done analytically);
\item a three-dimensional (photodetector plane location $+$ time of hit) Gaussian function is centered at the hit location and time of each MC photon;
\item the (unnormalized) PDF is the sum of these Gaussian functions.  
\end{itemize}
The procedure is repeated to construct the PDF for each mass hypothesis.  Since this approach naturally includes timing information, it removes many ambiguities in the hit patterns on the photodetector plane.   The so-called chromatic effect ($\lambda$ dependence of $\theta_C$) is accounted for as well as it can be, given the timing resolution of the photosensors (see Sec.~\ref{sec:pmt}), but the chromatic error is still one of the leading contributions to the per-photon $\theta_C$ resolution.  Variations on the optical box design were evaluated using this procedure, and the expected performance of the complete design is shown in Fig.~\ref{fig:performance}.

\subsection{Photon Detection and Readout}
\label{sec:pmt}

The imaging of Cherenkov photons for the~\gx{} DIRC detector requires a two-dimensional photodetector with good position resolution and moderate timing resolution ($\sim1$ ns) which can operate (with appropriate shielding) in the fringe magnetic field from the \gx{} solenoid.  Given these requirements, the Hamamatsu H12700 64-channel MaPMT photodetector will be used for the \gx{} DIRC, which will require 216 MaPMTs with a total of 13824 channels.  One advantage of this choice is a synergy with the RICH detector for CLAS12~\cite{rich_clas12} (also at Jefferson Lab), which also uses the H12700 and has similar requirements for the readout system.  Therefore, the \gx{} DIRC will utilize the same electronics as the CLAS12 RICH, which is based on the MAROC3 chip~\cite{maroc3} that is specifically designed for the readout of 64-channel MaPMTs.  

\begin{figure}
    \begin{center}
        \includegraphics[width=0.8\textwidth]{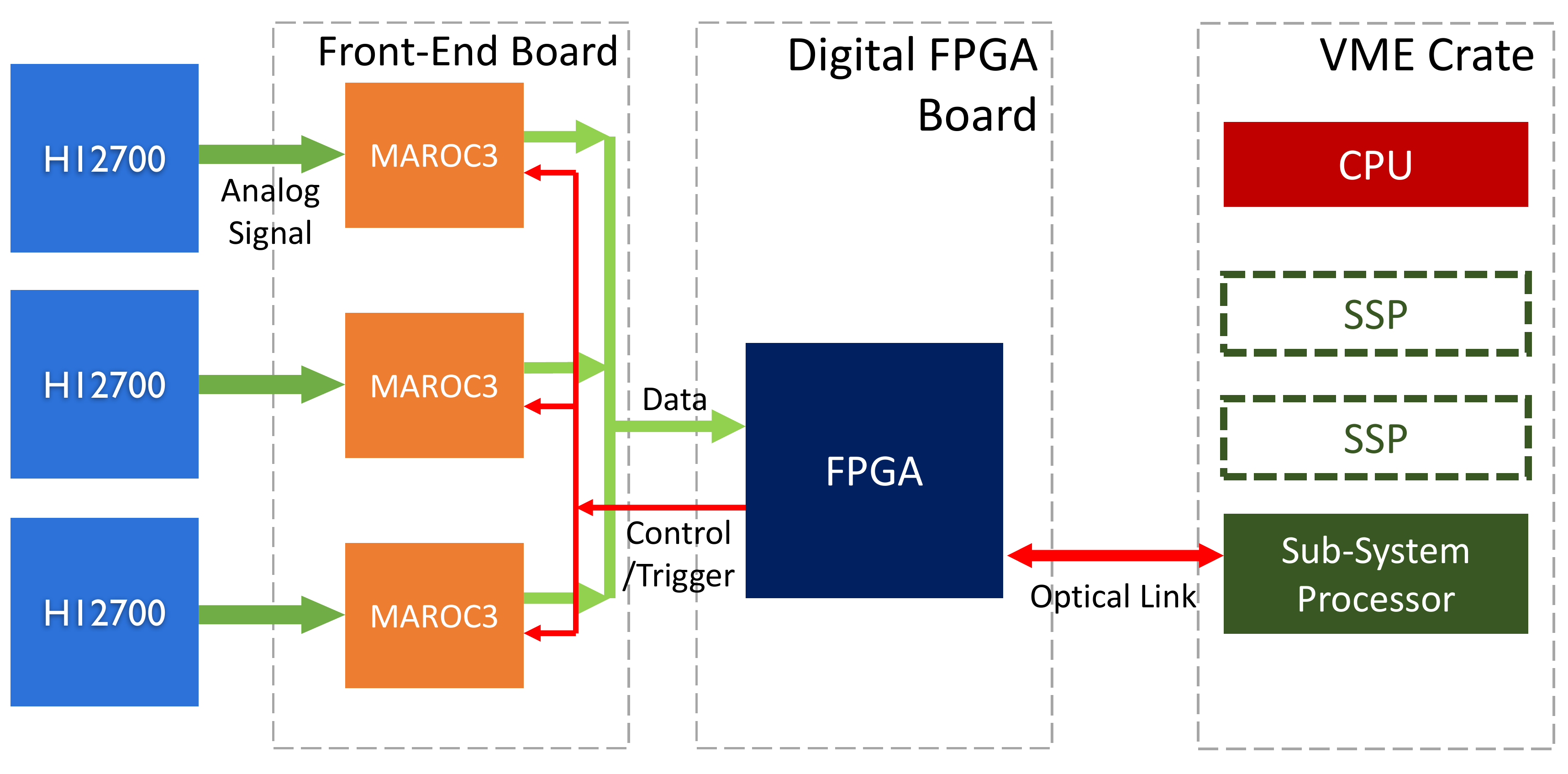}
    \caption{Readout scheme of the CLAS12 RICH and \gx{} DIRC.}\label{fig:mapmt_readout}
\end{center}
\end{figure}

As shown in the diagram of the readout scheme in Figure~\ref{fig:mapmt_readout}, the MAROC3 chips digitize the analog signals from the MaPMTs and pass the resulting binary data stream to a digital FPGA board.  The FPGA not only processes the data on board, but also controls and provides triggers to the MAROC3 chips.  The processed data from the FPGA will then be transmitted to a Jefferson Lab developed Sub-System Processor (SSP) hosted in a VME crate through high speed optical links, which seamlessly integrates into the current Hall D data acquisition system.  The front end (ASIC) board has already been developed by the CLAS12 RICH group at INFN, and the digital FPGA board, already in production, was developed by the Jefferson Lab fast electronics group.  

\section{Summary and Outlook}
\label{Sec:Summary}

The ability to reconstruct kaon final states is absolutely critical to maximizing the physics potential of the \gx{} experiment.  To enhance the particle identification capabilities of \gx{} a DIRC detector has been designed, utilizing decommissioned components of the BaBar DIRC.  The design for two new expansion volumes are being finalized, and construction is expected to begin in 2016.  Preparations for transporting the bar boxes from SLAC to Jefferson Lab are underway, with an initial installation of two bar boxes and one optical box in Hall D planned for 2017.  After this initial installation the performance of the system will be evaluated with GlueX beam data, with the final installation to follow in 2018.

\acknowledgments

This material is based upon work supported by the U.S. Department of Energy, Office of Science, Office of Nuclear Physics under contracts DE-AC05-06OR23177, DE-FG02-05ER41374 and DE-SC0010497.


\begin{thebibliography}{1}

  \bibitem{PAC30}
  \gx~Collaboration, ``Mapping the spectrum of light quark Mesons and gluonic excitations with linearly polarized protons," Jefferson Lab PAC 30 Proposal (2006), Available at: \href{http://www.gluex.org/docs/pac30 proposal.pdf}{\it http://www.gluex.org/docs/pac30 proposal.pdf}
  
  \bibitem{PAC40} 
  A.~AlekSejevs {\it et al.} (GlueX Collaboration),
  ``An initial study of mesons and baryons containing strange quarks with GlueX,'' Jefferson Lab PAC 40 Proposal (2013), arXiv:1305.1523 [nucl-ex].
 
 \bibitem{PAC42}
  M.~Dugger {\it et al.} (GlueX Collaboration),
  ``A study of decays to strange final states with GlueX in Hall D using components of the BaBar DIRC,'' Jefferson Lab PAC 42 Proposal (2014) arXiv:1408.0215 [physics.ins-det]. 

 \bibitem{Dudek:2013yja} 
  J.~J.~Dudek {\it et al.},
  ``Toward the excited isoscalar meson spectrum from lattice QCD,''
  Phys.\ Rev.\ D {\bf 88}, 094505 (2013), arXiv:1309.2608 [hep-lat].

 \bibitem{Ghoul:2015ifw} 
  H.~Al Ghoul {\it et al.} (GlueX Collaboration),
  ``First Results from The GlueX Experiment,''
  AIP Conf.\ Proc.\  {\bf 1735}, 020001 (2016), arXiv:1512.03699 [nucl-ex].
  
  \bibitem{dirc_tdr}
  \gx~Collaboration, ``GlueX DIRC Technical Design Report," (2015), Available at: \href{ http://argus.phys.uregina.ca/gluex/DocDB/0028/002809/003/dirc_tdr.pdf}{\it http://argus.phys.uregina.ca/gluex/DocDB/0028/002809/003/dirc tdr.pdf}
  
  \bibitem{Adam:2004fq} 
  I.~Adam {\it et al.} (BaBar DIRC Collaboration),
  ``The DIRC particle identification system for the BaBar experiment,''
  Nucl.\ Instrum.\ Meth.\ A {\bf 538}, 281 (2005).
  
  \bibitem{Dey:2014fda} 
  B.~Dey {\it et al.},
  ``Design and performance of the Focusing DIRC detector,''
  Nucl.\ Instrum.\ Meth.\ A {\bf 775}, 112 (2015), arXiv:1410.0075 [physics.ins-det].
 
 \bibitem{Dzhygadlo:2016uqi} 
  R.~Dzhygadlo {\it et al.},
  ``The PANDA Barrel DIRC,''
  JINST {\bf 11}, no. 05, C05013 (2016).
 
 \bibitem{rich_clas12}
 M.~Contalbrigo, E.~Cisbani, P.~Rossi, ``The CLAS12 large area RICH detector", Nucl. Instr. and Meth. {\bf A639}, 302 (2011).
  
 \bibitem{maroc3}
 S.~Blin {\it et al.}, ``A generic photomultiplier readout chip", JINST {\bf 5}, C12007 (2010).
  
\end{thebibliography}
\end{document}